# Above room temperature skyrmions in magnetic insulators

Qiming Shao[1*], Yawen Liu[2*], Guoqiang Yu[1,5], Se Kwon Kim[3,6], Xiaoyu Che[1], Chi Tang[2], Qing Lin He[1,7], Yaroslav Tserkovnyak[3], Jing Shi[2] and Kang L. Wang[1,3,4]

[1]Department of Electrical and Computer Engineering, University of California, Los Angeles, CA 90095, USA.

[2]Department of Physics and Astronomy, University of California, Riverside, CA 92521, USA.

[3]Department of Physics and Astronomy, University of California, Los Angeles, CA 90095, USA.

[4]Department of Materials Science and Engineering, University of California, Los Angeles, CA 90095, USA.

[5]Beijing National Laboratory for Condensed Matter Physics, Institute of Physics, Chinese Academy of Sciences, Beijing 100190, China.

[6] Department of Physics and Astronomy, University of Missouri, Columbia, Columbia, MO 65211, USA.

[7] International Center for Quantum Materials, School of Physics, Peking University, Beijing, 100871, China.

[*]These authors contributed to this work equally.

**Non-volatile memory and computing technology rely on efficient read and write of ultra-tiny information carriers that do not wear out. Magnetic skyrmions are emerging as a potential carrier since they are topologically robust nanoscale spin textures that can be manipulated with ultralow current density[1,2]. To data, most of skyrmions are reported in metallic films[3-10], which suffer from additional Ohmic loss and thus high energy dissipation. Therefore, skyrmions in magnetic insulators are of technological importance for low-power information processing applications due to their low damping and the absence of Ohmic loss. Moreover, they attract fundamental interest in studying various magnon-skyrmion interactions[11]. Skyrmions have been observed in one insulating material $Cu_2OSeO_3$ at cryogenic temperatures, where they are stabilized by bulk Dzyaloshinskii-Moriya interaction[12]. Here, we report the observation of magnetic skyrmions that survive above room temperature in magnetic insulator/heavy metal heterostructures, i.e., thulium iron garnet/platinum. The presence of these skyrmions results from the Dzyaloshinskii-Moriya interaction at the interface and is identified by the emergent topological Hall effect. Through tuning the magnetic anisotropy via varying temperature, we observe skyrmions in a large window of external magnetic field and enhanced stability of skyrmions in the easy-plane anisotropy regime. Our results will help create a new platform for insulating skyrmion-based room temperature low dissipation spintronic applications.**

More than four decades ago, movable magnetic bubbles in garnets and ferrites had excited huge interest for "magnetic bubble memory" applications [13]. However, two critical shortcomings precluded the commercialization of these bubble devices. First, the size of these bubbles was around 0.1 – 10 micrometers, which was too large for practical applications. Second, the manipulation of these bubbles required an on-chip magnetic field generator, which added significant complexity of circuit design and cost to the devices, making scaling difficult. The recently discovered skyrmions in B20 compounds and transition metal/ heavy metal thin films may easily overcome these two disadvantages and thus again ignite the interest of using skyrmions as information carriers [1,3-10,12]. First, the size of skyrmions has been scaled down to sub-100 nanometers in material systems that have appreciable Dzyaloshinskii-Moriya interaction (DMI) due to inversion symmetry breaking either in bulk or at the interface [4]. Second, skyrmions can be moved by using low threshold electric current and by electric field [2,7,14], which makes the scaling of skyrmion-based devices much more convenient compared with the case of using external magnetic field. Furthermore, for memory applications, the writing of skyrmions using spin-polarized current has been demonstrated at room temperature [6,10], and the electrical detection (reading) of skyrmions can be achieved with the topological Hall effect (THE); THE is resulted from the Berry phase acquired by the spin-polarized carriers going through a skyrmion texture [15,16].

Magnetic insulators that host skyrmions are particularly attractive since they have very low damping and allow long-distance information transmission free of Joule heating [17]. Moreover, various exotic phenomena based on magnon-skyrmion interactions [11], like magnon quantum Hall [18], long-range magnon transport [19], and magnon driven skyrmion motion [20], have been predicted in insulating skyrmion systems. To date, however, the only B20 magnetic insulator ($Cu_2OSeO_3$) that has been reported to host bulk DMI-stabilized skyrmions has a Curie temperature ($T_C$ ~ 60 K) [12]. In this insulating skyrmion system, magnetic excitations[21] and thermally-driven skyrmion motion[22,23] have been observed. The choice of magnetic insulators is limited due to the strict requirement of the crystal structure with inversion symmetry breaking, which is essential to generate strong DMI. The commonly studied high-temperature magnetic insulators, like garnets and ferrites, are centrosymmetric and thus magnetic bubbles lack a preferred chirality due to the absence of DMI [24].

In this Letter, we demonstrate the electrical detection of above-room-temperature magnetic skyrmions using a pronounced THE in a bilayer heterostructure composed of a magnetic insulator thulium iron garnet ($Tm_3Fe_5O_{12}$, TmIG) thin film in contact with a Pt film. The $T_C$ for bulk TmIG is ~ 560 K [13]. The skyrmions are stabilized by the interfacial DMI, which is the result of strong spin-orbit coupling and inversion symmetry breaking at the TmIG/Pt interface. The THE is enabled by the exchange coupling between the skyrmions in TmIG and the finite spin polarization in the bottom of Pt layer (Fig. 1a). By varying the temperature, the magnetic anisotropy of TmIG can be tuned from easy axis anisotropy (out-of-film-plane) to easy plane anisotropy (in-the-film-plane), which allows for investigating the stability of skyrmions in both cases. Our experimentally observed skyrmion phase diagram established from the THE is consistent with the ones obtained by using analytical calculations and micromagnetic simulations. We discover an enhanced stability of skyrmions against the external field when the magnetic anisotropy is transitioned from the easy axis to the easy plane. At last, we show that the skyrmion phase diagram becomes smaller and eventually diminishes as the TmIG thickness increases, which is consistent with the interfacial DMI picture.

TmIG thin films are first grown by pulsed laser deposition on $Nd_3Ga_5O_{12}$ substrate. The robust perpendicular magnetic anisotropy (PMA) is obtained at room temperature through the strain-induced magneto-elastic effect as a result of the lattice mismatch between TmIG and $Nd_3Ga_5O_{12}$ [25]. Then, a thin 4 nm-thick Pt layer is sputtered on the TmIG at room temperature. The exchange coupling between TmIG and Pt results in the anomalous Hall effect (AHE) and spin Hall magnetoresistance (SMR) at and above room temperature in a patterned Hall bar device [25,26].

Assuming a smooth spin texture, we have a generic expression for antisymmetric Hall resistivity ($\rho_{xy}$) obtained on symmetry grounds (see Supplementary Information S1)

$$\rho_{xy} = \rho_o B_z + \rho_A m_z + \frac{\rho_T}{4\pi} \iint d^2\mathbf{r}\, \mathbf{m} \cdot \left(\frac{\partial \mathbf{m}}{\partial x} \times \frac{\partial \mathbf{m}}{\partial y}\right), \qquad (1)$$

where $\rho_o$ is ordinary Hall effect (OHE) coefficient, $\rho_A$ is the saturation AHE resistivity, $m_z$ is the average z-component of magnetization unit vector in the Hall contact area and the third term is the topological Hall effect (THE) contribution ($\rho_{THE}$). In the THE term, $\rho_T$ is the THE coefficient and the integral counts how many times $\mathbf{m}(\mathbf{r}) = \mathbf{m}(x, y)$ wraps a unit sphere, which is the skyrmion number in real space.

We observe a typical sharp hysteresis loop of $\rho_{xy}$ as a function of out-of-plane external field ($B_Z$) for the TmIG (3.2 nm)/Pt(4 nm) bilayer at 350 K (Fig. 1b), where the step function at low fields is due to the AHE and the linear background with a negative slope at large fields arises from the OHE. Above 350 K, unusual $\rho_{xy}$ dips at low positive fields and peaks at low negative fields emerge and gradually disappear at large fields as shown in Fig. 1b. We identity the overshoot in these out-of-plane hysteresis loops as the THE due to the presence of magnetic skyrmions [27,28]. To obtain the AHE contribution in $\rho_{xy}$, we determine the $m_z$ as a function of $B_z$ by tracking the change of longitudinal resistance $\Delta\rho_{yy}$ (Fig. 1c) since $\Delta\rho_{yy} \propto m_z^2$ according to the theory of SMR [26] (see Supplementary Information S2). We plot the measured $\rho_{xy}$ and the simulated contributions from the OHE and AHE together in Fig. 2a for $T = 360$ K, where we observe an apparent difference between these two plots. By subtracting the contributions from the OHE and AHE, we determine the magnitude of the $\rho_{THE}$ (Fig. 2a).

To host magnetic skyrmions in the TmIG, there must be a sizable interfacial DMI energy ($D$) at the interface between the TmIG and the Pt for stabilizing magnetic chiral structures. Experimentally, Pt/ferromagnetic metal bilayers have been reported to show a very strong interfacial DMI, $D \sim 1\text{-}2$ mJ/m$^2$, which supports sub-100 nm skyrmions at room temperature [7,8]. In theory, we also expect to have a sizeable interfacial DMI at the TmIG/Pt interface due to a strong coupling between Pt and Fe atoms as evidenced by the AHE and SMR. We estimate the magnitude of $D$ in our TmIG/Pt bilayer by employing a domain wall motion technique described in ref. [29]. The determined $D$ is $\sim 51$ µJ/m$^2$ at room temperature (see Supplementary Information S3). While the absolute value of $D$ at room temperature for our TmIG/Pt is smaller than the case in Pt/ferromagnetic metal bilayers, the ratio of $D$ over exchange stiffness ($A$) is comparable since $A$ is estimated to be $\sim 0.84$ pJ/m (see Supplementary Information S4). As the temperature increases, both $D$ and $A$ decrease as the saturation magnetization ($M_S$) decreases. The $D$ and $A$ have the linear relation according to ref. [30] for the reason that both are rooted to the same exchange mechanism. Therefore, their ratio $D/A$ remains almost constant as the temperature varies.

The appropriate anisotropy energy ($K$) of TmIG is achieved by varying the temperature to satisfy the requirement for the presence of skyrmions. Theory shows that $K$ should be $\leq \frac{\pi^2}{16}\frac{D^2}{A}$ to form a skyrmion lattice [31,32], which suggests that skyrmions can only be stabilized in a weak anisotropy regime. For our TmIG/Pt, the $K$ can be continuously tuned from positive (PMA) to negative (easy plane anisotropy) by changing the temperature, which passes the zero-anisotropy energy. This tunability is critical for the formation of skyrmions in a window of temperature. The increase of temperature reduces $K$, most likely due to the reduced magneto-elastic coefficient that contributes to the PMA [33]. By varying the temperature, we obtain a skyrmion phase diagram from the THE as a function of temperature and external field (Fig. 2b). To illustrate the importance of the tunable $K$, we plot the temperature dependence of $K$ as shown in Fig. 2c. Below 300 K, the $K$ is much larger than $\frac{\pi^2}{16}\frac{D^2}{A}$ and thus the ferromagnetic phase is always favored. As the temperature increases above 345 K, $K$ reduces to a value smaller than $\frac{\pi^2}{16}\frac{D^2}{A}$, where skyrmions could emerge. In experiments, the THE occurs above 350 K, which agrees with the theory. With the sizable $D$ and appropriate $K$, we conclude that the presence of magnetic skyrmions is the driving force for the observation of THE in the TmIG/Pt.

We now focus on the external field dependence of THE. We anticipate a spin spiral phase (or a balanced number of skyrmions with topological charge +1 and −1) near zero field and a ferromagnetic phase at large fields, in which the THE is minimized [15]. We estimate the stability of a skyrmion lattice by employing a free energy minimization method, in which we consider exchange, anisotropy, DMI, and Zeeman energy [31,34]. Here, we assume a perfect hexagonal skyrmion lattice for simplicity of calculation. Since the $\rho_{THE}$ is proportional to the skyrmion density, we compute a skyrmion density diagram as a function of normalized $K$ and Zeeman energy (Fig. 2d), where $B_E = M_S B_Z$ (see Supplementary Information S5). Full micromagnetic simulations reveal an even larger skyrmion window of $K$ and $B_Z$ when the magnetostatic energy is included (see Supplementary Information S6). In agreement with the calculated skyrmion density in Fig. 2d, the $\rho_{THE}$ at a given temperature first increases and then decreases with the external field (Fig. 2b). Also, below 370 K, the $\rho_{THE}$ at a given field increases as temperature increases (Fig. 2b) due to the reduced PMA (Fig. 2c), which agrees with Fig. 2d and is consistent with the very recent observation in Ir/Fe/Co/Pt multilayers [32]. Furthermore, we observe a larger external field window in a higher temperature for stable skyrmions in Fig. 2b, when the $K$ transitions from PMA ($K > 0$) to easy plane anisotropy ($K < 0$) near 370 K (Fig. 2c). Thanks to the great tunablity of $K$ in the TmIG/Pt bilayer through varying temperature, the stability of skyrmions against external field is enhanced and the $\rho_{THE}$ is increased in the easy plane anisotropy regime. Our observations are consistent with the calculations (Fig. 2d) and the prediction by Banerjee *et al*.[34]. Therefore, the temperature and external field dependences of THE agree with the theoretical expectations, confirming the existence of magnetic skyrmions in TmIG/Pt.

We now investigate the TmIG thickness ($t_{TmIG}$) dependent THE in TmIG/Pt(4 nm) bilayers to clarify the importance of the interfacial DMI. Since the magnitude of interfacial DMI is inversely proportional to the ferromagnetic layer thickness [35], we do not expect the presence of magnetic skyrmions in the thick TmIG limit. In the extremely thin TmIG, the ferromagnetism disappears. As the $t_{TmIG}$ increases, the relative strength of the interfacial DMI to the exchange stiffness ($D/A$) reduces, resulting in a larger skyrmion size. In addition, since the $D^2/A$ decreases with the increasing $t_{TmIG}$, the appropriate range of $K$ and $B_Z$ for stabilizing skyrmions shrinks, leading to a smaller skyrmion phase diagram with respect to temperature and external field. Experimentally, the results of skyrmion phase diagram from THE from the $t_{TmIG}$ = 3.2 nm, 4 nm and 6 nm are shown in Figs. 3A-C, respectively. We have two major findings. First, the highest $\rho_{THE}$ reduces from 4.58 to 1.66 nΩ·cm when the $t_{TmIG}$ increases from 3.2 nm to 6 nm. The highest $\rho_{THE}$ is an indicator of skyrmion size (density); the larger the $\rho_{THE}$ is, the smaller (denser) the skyrmion is. This agrees with the decreasing $D/A$ as the $t_{TmIG}$ increases. Second, the temperature-field window of skyrmions becomes smaller, which is consistent with the interfacial DMI picture as described above. For the $t_{TmIG}$ = 6.4 nm, we do not observe a clear THE signal (see Supplementary Information S7).

In addition to the electrical THE detection of skyrmions, the current-induced spin-orbit torques in the magnetic insulator/heavy metal bilayer could provide an efficient way to manipulate the skyrmion in the magnetic insulator. We have shown that the strong spin-orbit torque generated by the current in Pt can efficiently switch the magnetization of TmIG at room temperature (see Supplementary Information S8). Therefore, we believe that the discovery of skyrmions in a simple magnetic insulator/heavy metal bilayer heterostructure like TmIG/Pt encourages and promises enormous future efforts for realizing low-power skyrmion-based applications at room temperature beyond studying fundamental problems such as magnon-skyrmion interaction in magnetic insulators. There are also many open questions remaining. For example, current-driven skyrmion dynamics, direct imaging of a skyrmion, and behaviors of the magnon-skyrmion interaction in the magnetic insulator/heavy metal bilayer system require further investigation.

**Methods**

**Material growth and preparation.** Deposition of high-quality ferrimagnetic insulator TmIG with PMA on substituted gadolinium gallium garnet (SGGG) substrate with pulse laser deposition has been demonstrated in previous work[25]. Here we choose $Nd_3Ga_5O_{12}$ (NGG) as the substrate, which has a very close lattice constant to SGGG. TmIG films are grown at a low temperature of about 200 ℃ and an oxygen pressure of 0.3 mTorr, and post annealed at 850℃ for 200s with sufficient oxygen gas flow. TmIG thickness is determined using a pre-calibrated growth rate. Strong PMA is confirmed by perpendicular magnetization measurement with vibrating sample magnetometer. Our atomic force microscope image indicates a high-quality atomic flat surface with a root-mean-square roughness ~ 0.14 nm. After careful characterizations, NGG/TmIG thin films are transferred to a high-vacuum magnetron sputtering chamber. We perform a light Ar plasma cleaning of the TmIG surface first and then a thin polycrystalline layer of Pt is deposited.

**Device fabrication and electrical measurement.** The whole TmIG/Pt films are patterned into Hall bar structures with a channel width of 20 µm by using standard photo-lithography and dry etching. Then, contact metals Cr(10 nm)/Au(100 nm) are deposited using e-beam evaporation. The electrical measurement is performed using lock-in technique in a physical property measurement system.

**Acknowledgements** We thank Junxue Li for help on thin film preparation, Cheng Zheng and Aryan Navabi for assistance of device fabrication and Yizhou Liu for helpful discussions on micromagnetic simulations. Qiming Shao thanks Peng Zhang for assistance of loop shift measurements. This work is supported partially by the Spins and Heat in Nanoscale Electronic Systems (SHINES), an Energy Frontier Research Center funded by the US Department of Energy (DOE), Office of Science, Basic Energy Sciences (BES), under Award # DE-SC0012670. We acknowledge the support from the Army Research Office Multidisciplinary University Research Initiative (MURI) program accomplished under Grant Number W911NF-16-1-0472 and W911NF-15-1-10561. The authors at UCLA are also partially supported by the National Science Foundation (ECCS 1611570), and by C-SPIN and FAME, two of six centers of STARnet, a Semiconductor Research Corporation program, sponsored by MARCO and DARPA. Se Kwon Kim is supported by the Army Research Office under Contract No. W911NF-14-1-0016.

**Author information** Correspondence and requests for materials should be addressed to K.L.W. (wang@seas.ucla.edu).

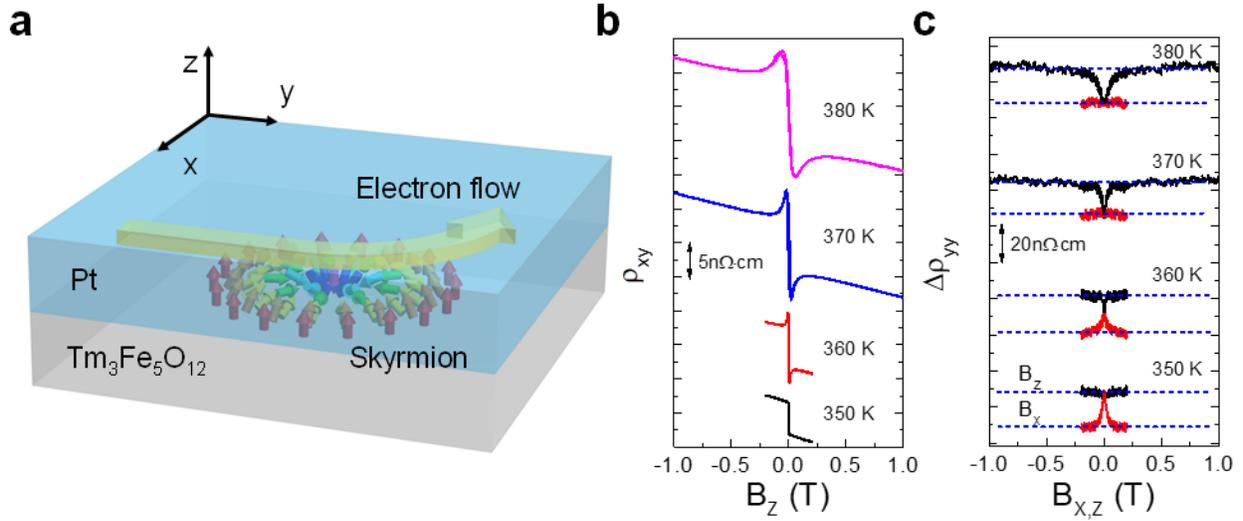

**Figure 1. Illustration of topological Hall effect in the TmIG/Pt and transport properties of the TmIG(3.2 nm)/Pt (4 nm) bilayer. a,** Schematic of the topological Hall effect in the TmIG/Pt. The current at the TmIG/Pt interface goes through the emergent electromagnetic field generated by the skyrmion in the TmIG and gives rise to the transverse Hall current. **b,** Hall resistivity as a function of the out-of-plane magnetic field at different temperatures. Above 350 K, topological Hall effect is observed as peaks and dips happen at low fields. **c,** Longitudinal resistivity as a function of both the out-of-plane (black, along the $\pm z$ direction) and in-plane (red, along the $\pm x$ direction) magnetic fields at different temperatures, from which we determine the out-of-plane magnetization component of TmIG as a function of external field using the theory of spin Hall magnetoresistance (see Supplementary Information S2). The data are offset for clarity.

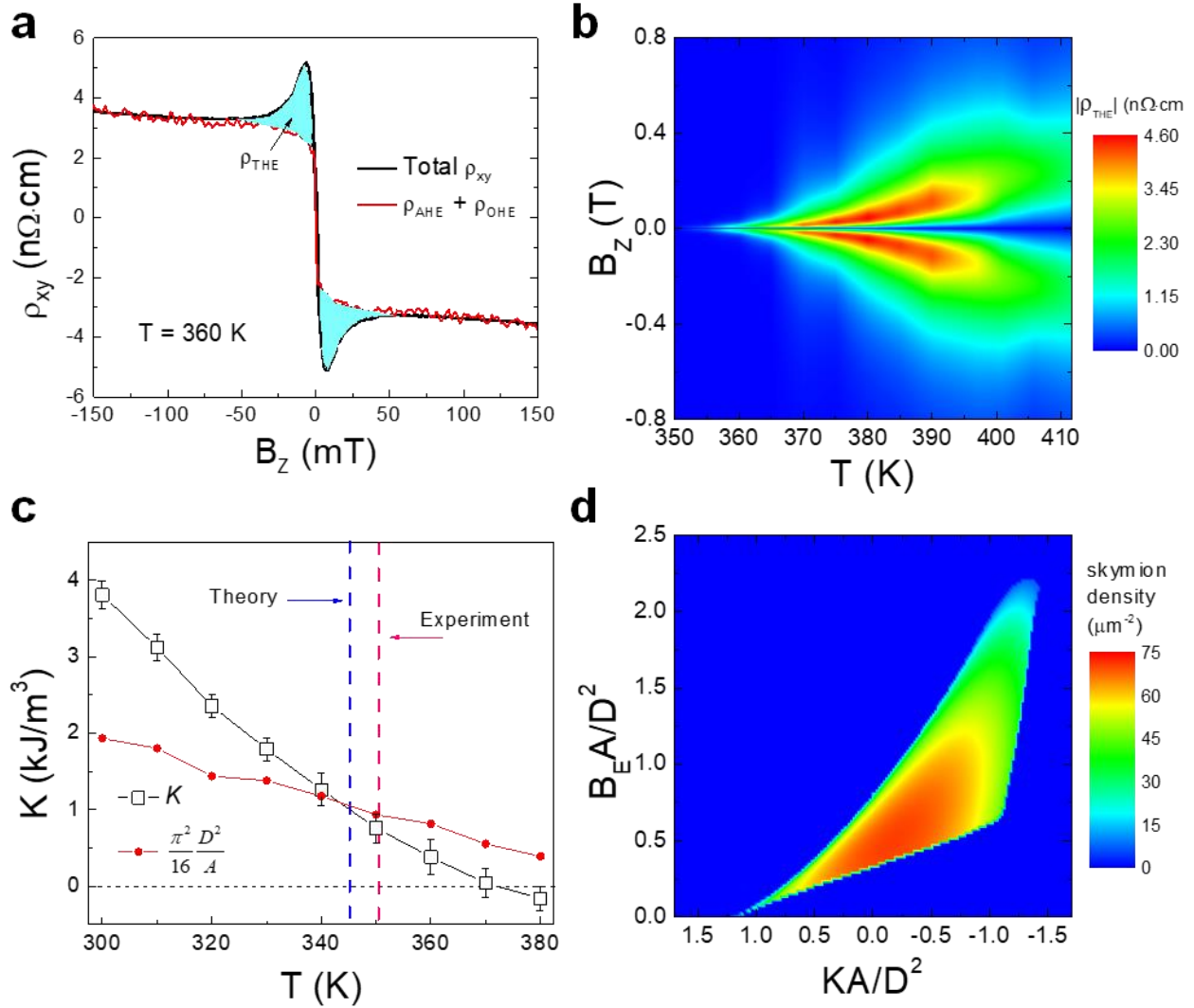

**Figure 2. Observation of topological Hall effect (THE) in the TmIG(3.2 nm)/Pt (4 nm) bilayer. a,** Hall resistivity (black curve) as a function of an out-of-plane magnetic field at $T = 360$ K. The red curve is the contribution of the anomalous Hall effect and the ordinary Hall effect. The shaded light blue region is the contribution of THE. **b,** Skyrmion phase diagram from the THE as a function of temperature $T$ and external field $B_Z$. The color bar indicates the value of measured THE resistivity. **c,** Anisotropy energy $K$ (black square symbols and curve) as a function of $T$. The red circle symbols and curve show the boundary of $K$, below which magnetic skyrmions exist [31]. Here, $D$ is interfacial DMI energy and $A$ is the exchange stiffness. Following the trace of $T$-dependent $K$, theory predicts that magnetic skyrmions appear above 345 K, which is in a good agreement with the experimental value 350 K. The anisotropy transitions from perpendicular magnetic anisotropy to easy-plane anisotropy near 370 K. **d,** Theoretical skyrmion density diagram as a function of the normalized anisotropy energy ($KA/D^2$) and the Zeeman energy ($B_EA/D^2$).

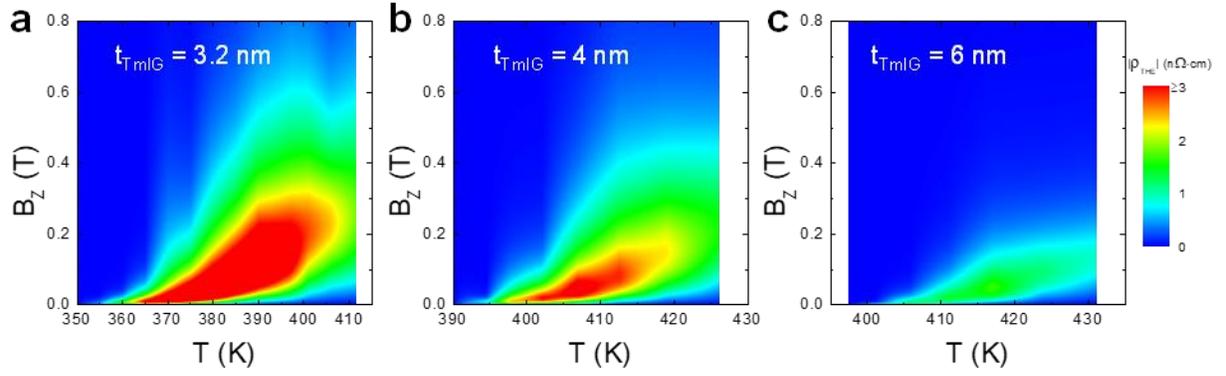

**Figure 3. Experimentally obtained evolution of skyrmion phase diagram as a function of TmIG thickness ($t_{TmIG}$) in TmIG/Pt (4 nm) bilayers. a,** $t_{TmIG}$ = 3.2 nm. **b,** $t_{TmIG}$ = 4 nm. **c,** $t_{TmIG}$ = 6 nm. Note that the highest THE resistivity drops as the $t_{TmIG}$ increases, which are 4.58 nΩ·cm, 3.44 nΩ·cm, 1.66 nΩ·cm for the 3.2 nm, 4 nm and 6 nm, respectively (see Fig. 2b and Supplementary Information S7). We set the highest value of the plotted THE resistivity to 3 nΩ·cm to show a good color contrast (see color bar).

17